\documentclass[aps,prl,superscriptaddress,twocolumn,nopacs,amsmath,amssymb]{revtex4}

\usepackage{color}
\usepackage{graphicx}
\usepackage{bm}
\usepackage{amssymb, amsmath, amsbsy}

\begin{document}

\title{Bending modes, anharmonic effects and thermal expansion coefficient in single layer and multilayer graphene
}

\author{P. L. de Andres}
\author{F. Guinea}

\affiliation{
Instituto de Ciencia de Materiales de Madrid (CSIC),
Cantoblanco, 28049 Madrid, Spain.
}
\author{M. I. Katsnelson}
\affiliation{
Radboud University Nijmegen, Institute for Molecules and Materials,
Heyendaalseweg 135, 6525AJ Nijmegen, The Netherlands.
}

\date{\today}

\begin{abstract}
We present a simple analytical approach to study anharmonic effects in single layer, bilayer, and multilayer graphene. The coupling between in plane and out of plane modes leads to negative Gr\"uneisen coefficients and negative thermal expansion. The value of the thermal expansion coefficient depends on the coupling to the substrate. The bending rigidity in bilayer graphene shows a crossover between a long wavelength regime where its value is determined by the in plane elastic properties and a short wavelength regime where its value approaches twice that of a single layer.
\end{abstract}


% 63.22.Rc Phonons in graphene.
% 65.80.Ck Thermal properties of graphene.
% 61.48.Gh Structure of graphene.

% 62.20.-x Mechanical properties of solids.


%Use showkeys class option if keyword display desired

\maketitle

{\em Introduction.}

To realize the full potential of graphene layers in promising applications,
like the design of fast
electronic devices or sensitive and accurate molecular detectors,
it is important to reach a thorough understanding of its properties
down to the atomic level.
\cite{novoselov04,schedin07}
At T= 0 K, in the absence of defects,
the carbon bond on the graphene layer is well understood in terms of
the formation of three in-plane localized strong sp$^{2}$ bonds, and
a fourth delocalized, out-of-plane, $\pi$-like bond.\cite{NGPNG09,K12}
The optimum geometrical configuration is achieved by
a honeycomb lattice formed by two equivalent sublattices
displaying P6/mmm symmetry.
The corresponding electronic structure shows bands
dispersing linearly around the Fermi energy that are
responsible for the fast and efficient transport of carriers.
Both experimentally and theoretically,\cite{lee08,Mounet05,ZKF09} it is
shown that this kind of arrangement results in a material with the largest
in-plane elastic constants known yet.
Therefore the 2D perfect flat layer makes the most stable configuration since
deviations from a common plane requires a significant amount of energy.
Any departure from such a scenario affects greatly the atomic scale
properties of the layer and must be understood in order to efficiently exploit
graphene's properties.
Different reasons, however, might be invoked for a two-dimensional graphene layer to adopt a
certain corrugation at different scales. First, at a non-zero temperature
a thermodynamic argument implies the impossibility for a perfect 2D layer to exist
in 3D.\cite{landauSP1,NP87,AL88,doussal92,FLK07,K12}
Second, defects like adsorbed impurities, vacancies, etc,
create local corrugations at the atomic scale\cite{BKL08} that propagate via the
elastic properties of the lattice
originating long-range correlations.
Finally, external applied stresses related to conditions on the boundary
make graphene to bend and to corrugate;
an interesting point to study since the growth of graphene layers on different
supporting substrates implies mismatches that introduce
all kind of stresses that have been observed to originate
a highly complex and corrugated landscape.\cite{morgenstern10,KCNG11}
In this work, we analyze a simple model based on the
theory of elasticity to obtain physical insight on the
Gr\"uneisen coefficients and the thermal expansion coefficient of graphene,\cite{lau09}
which can be compared to atomistic models based on {\it ab initio}
Density Functional Theory that yields a realistic quantitative description of
bending modes and corrugations appearing at the atomic scale\cite{AGK12}.

{\em Single layer graphene.}

We study anharmonic effects using the continuum theory of elasticity. We extend previous analyses\cite{L52,A01}, using the standard theory of free standing membranes\cite{NP87,AL88,doussal92,K12}.  The Hamiltonian is\cite{landauEl}
%\begin{widetext}
\begin{align}
{\cal H}_{el} &=  \int d^2 \vec{\bf r} \left\{ \frac{\rho}{2} \left[ \left( \partial_t u_x \right)^2 + \left( \partial_t u_y \right)^2 + \left( \partial_t h \right)^2 \right]  \right. + \nonumber \\ &+ \frac{\kappa}{2} \left( \frac{\partial^2 h}{\partial x^2} + \frac{\partial^2 h}{\partial y^2} \right)^2 + \nonumber \\ &+ \frac{\lambda}{2}   \left(  \partial_x u_x + \partial_y u_y + \frac{\left( \partial_x h \right)^2 + \left( \partial_y h \right)^2}{2}  \right)^2 + \nonumber \\ &+ \mu \left[ \left( \partial_x u_x + \frac{\left( \partial_x h \right)^2}{2} \right)^2 + \left( \partial_y u_y + \frac{\left( \partial_y h \right)^2}{2} \right)^2 +  \right. \nonumber \\ &+ \left. \left. \frac{\left[ \partial_x u_y + \partial_y u_x + \left( \partial_x h \right) \left( \partial_y h \right) \right]^2}{2} \right] \right\}
\label{hamil}
\end{align}
%\end{widetext}
where $\rho$ is the mass density, $\vec{\bf u}$ is the two dimensional displacement vector, $h$ is the displacement in the out of plane direction, $\kappa$ is the bending rigidity, and $\lambda$ and $\mu$ are elastic Lam\'e coefficients. For graphene, we have\cite{ZKF09} $\kappa \approx 1$ eV, $\lambda = 2$ eV \AA$^{-2}$ and $\mu = 10$ eV \AA$^{-2}$.

We study the modes associated to the out of plane displacements. If we assume that there are no in plane tensions, $\partial_i u_j = 0$, and we neglect the quartic terms in $h$, we obtain $\omega_{\vec{\bf q}} = \sqrt{\kappa \left| \vec{\bf q} \right|^4 / \rho}$.
%\begin{align}
% \omega_{\vec{\bf q}} &= \sqrt{\frac{\kappa \left| \vec{\bf q} \right|^4}{\rho}}
% \label{frequency}
%\end{align}
This is the well known dispersion relation for out of plane flexural modes. We now analyze how these frequencies are modified when the in plane lattice constant is modified. An isotropic change of the lattice constant by a factor $\bar u$ can be included in the Hamiltonian, eq.~\ref{hamil}, by assuming that $\partial_x u_x = \partial_y u_y = \bar{u}$. The effective Hamiltonian for $h$, expanded to second order, becomes
%\begin{widetext}
\begin{align}
{\cal H}_{flex} &=  \int d^2 \vec{\bf r} \left\{ \frac{\rho}{2} \left( \partial_t h \right)^2 + \frac{\kappa}{2} \left( \frac{\partial^2 h}{\partial x^2} +  \frac{\partial^2 h}{\partial y^2} \right) + \right. \nonumber \\ &+ \left. \left( \lambda + \mu \right) \bar{u}  \left[ \left( \partial_x h \right)^2 + \left( \partial_y h \right)^2 \right] \right\}
\label{hamil11}
\end{align}
%\end{widetext}
The new frequencies of the flexural phonons are
\begin{align}
 \omega_{\vec{\bf q}} &= \sqrt{\frac{\kappa \left| \vec{\bf q} \right|^4 + 2\left( \lambda + \mu \right) \bar{u} \left| \vec{\bf q} \right|^2}{\rho}}
 \label{frequency}
\end{align}
The derivative of the phonon frequency with respect to a change in the area of the unit cell ${\cal A}$ is
\begin{align}
\gamma_{\vec{\bf q}} &= - \frac{{\cal A}}{\omega_{\vec{\bf q}}} \frac{\partial \omega_{\vec{\bf q}}}{\partial {\cal A}} = - \left. \frac{1}{2\omega_{\vec{\bf q}}} \frac{\partial \omega_{\vec{\bf q}}}{\partial \bar{u}} \right|_{\bar{u}=0} = - \frac{\lambda + \mu}{2 \kappa \left| \vec{\bf q} \right|^2}
\label{gruneisen}
\end{align}
where $\gamma_{\vec{\bf q}}$ is the Gr\"uneisen parameter. We obtain a negative Gr\"uneisen parameter for all low frequency flexural modes, which diverge for $\left| \vec{\bf q} \right| \rightarrow 0$ as $\left| \vec{\bf q} \right|^{-2}$. This expression is valid for momenta much smaller than the inverse of the interatomic spacing $a$, $\left| \vec{\bf q} \right| \ll a^{-1}$. This result is consistent with a number of numerical calculations, which show negative Gr\"uneisen coefficients for flexural modes, which tend to diverge at low momenta\cite{Mounet05,ZKF09,KF11,Petal11}.

Within the harmonic approximation, the estimate of the Gr\"uneisen parameters in eq.~\ref{gruneisen} allows us to obtain the thermal expansion coefficient \cite{K12}
\begin{align}
\alpha &= \frac{k_B}{{\cal A} \left( \lambda + \mu \right)} \sum_{\vec{\bf q}}  \left( \frac{\hbar \omega_{\vec{\bf q}}}{2 k_B T} \right)^2\frac{\gamma_{\vec{\bf q}}}{\sinh^2 \left( \frac{\hbar \omega_{\vec{\bf q}}}{2 k_B T} \right)}
\label{expansion}
\end{align}
where ${\cal A}$ is the area of the unit cell and we take into account that the two-dimensional bulk modulus $B = \lambda + \mu$. The sum (in thermodynamic limit is replaced by an integral) in the right-hand side of eq.~(\ref{expansion}) is divergent at small $q$ which is the consequence of inapplicability of the harmonic approximation at small $q$ where renormalization of effective bending rigidity and elastic modulii become relevant. The crossover wave vector is \cite{NP87,K12}
\begin{equation}
q^{\ast }=\sqrt{\frac{3k_{B}TY}{8\pi \kappa ^{2}}}  \label{qstar}
\end{equation}
where $Y =  4 \mu ( \lambda + \mu ) / ( \lambda + 2 \mu )$
%\begin{equation}
%Y=\frac{4\left( \lambda +\mu \right) \mu }{\left( \lambda +2\mu \right) }
%\label{Y}
%\end{equation}
is the two-dimensional Young modulus. Note that the corresponding phonon frequency lies deeply in the classical region:
\begin{equation}
\hbar \omega ^{\ast }=\frac{3k_{B}T}{8\pi }\frac{Y}{\sqrt{\kappa ^{3}\rho }}%
\sim k_{B}T\sqrt{\frac{m}{M}}\ll k_{B}T  \label{omegastar}
\end{equation}
where $m$ and $M$ are electron mass and mass of carbon atom, respectively. With the logarithmic accuracy,
\begin{equation}
\alpha \approx -\frac{k_{B}}{4\pi \kappa }\int\limits_{q^{\ast }}^{q_{T}}%
\frac{dq}{q}=-\frac{k_{B}}{8\pi \kappa }\ln \frac{k_{B}T}{\hbar \omega
^{\ast }}\approx -\frac{k_{B}}{16\pi \kappa }\ln \frac{\kappa ^{3}\rho }{%
\hbar ^{2}Y^{2}}  \label{alpha}
\end{equation}
where $q_T$ is the thermal wave vector satisfying the condition $\hbar \omega (q_T) = k_B T$. From the estimation in eq. (\ref{alpha}), we obtain $\alpha \approx - 10^{-5}$ K$^{-1}$, a quite good estimation for so oversimplified model (cf. Refs. \onlinecite{Mounet05,ZKF09}). Here we assume that the temperature is smaller than the maximal energy of the flexural phonon, $T_m \approx 15$ THz $\approx$ 700 K \cite{Mounet05}, otherwise one needs to add the factor $T_m /T$ under the argument of logarithm in eq. (\ref{alpha}).

Due to eq. (\ref{omegastar}) phonons relevant for the thermal expansion coefficient can be considered as classical at any temperatures. This allows us to repeat the calculation of $\alpha$ taking into account anharmonic effects. Due to eq. (\ref{hamil11}) and Hellmann-Feynman theorem the derivative of the free energy ${\cal F}$ with respect to the deformation at $\bar{u}=0$ can be rigorously expressed via the correlation function of out-of-plane displacements:
\begin{equation}
\frac{\partial \mathcal{F}}{\partial \overline{u}}=\left\langle \frac{%
\partial \mathcal{H}_{flex}}{\partial \overline{u}}\right\rangle =\left(
\lambda +\mu \right) \sum\limits_{\overrightarrow{q}}q^{2}\left\langle
\left\vert h_{\overrightarrow{q}}\right\vert ^{2}\right\rangle   \label{HF1}
\end{equation}
and via the anharmonic self energy $\Sigma \left( \overrightarrow{q}\right) $%
:%
\begin{equation}
\left\langle \left\vert h_{\overrightarrow{q}}\right\vert ^{2}\right\rangle =%
\frac{k_{B}T}{\kappa q^{4}+\Sigma \left( \overrightarrow{q}\right) }
\label{HF2}
\end{equation}%
The latter can be estimated from the condition that at $q=q^{\ast }$ both
terms in the denominator in eq. (\ref{HF2}) are of the same order of
magnitude \cite{K10}:%
\begin{equation}
\Sigma \left( q\right) =A\left( Yk_{B}T\right) ^{\eta /2}\kappa ^{1-\eta
}q^{4-\eta }  \label{sigma}
\end{equation}%
where $\eta \approx 0.85$ is the exponent of renormalization of the bending
rigidity; the numerical factor $A$ was calculated within the self-consistent
screening approximation \cite{RFZK11}; it was also shown that this
approximation agrees quite well with the atomistic Monte Carlo simulations.
Substituting eq. (\ref{sigma}) into eq. (\ref{HF2}) and further into  eq. (%
\ref{HF1}) one can calculate the thermal expansion coefficient%
\begin{equation}
\alpha =-\frac{1}{2\left( \lambda +\mu \right) }\frac{\partial ^{2}\mathcal{F%
}}{\partial T\partial \overline{u}}  \label{HF3}
\end{equation}%
with anharmonic effects taken into account. With the logarithmic accuracy,
the result coincides with eq. (\ref{alpha}). Thus, the contribution of
flexural mode to the thermal expansion coefficient is always negative and
temperature independent up to $T\approx T_{m}\approx 700$ K; at higher
temperatures it depends on the temperature logarithmically. This means that
the inversion of sign of the thermal expansion coefficient at high
temperature found in atomistic simulations \cite{ZKF09} is due to
contributions of other phonon modes.

This justifies the use of quasiharmonic approximation to estimate the
contribution of flexural phonons to the thermal expansion. Further we will
consider only this approximation.

Finally, from eq.~\ref{frequency} for the phonon frequencies we can estimate the momentum $q_c$ for which the value of $\omega_{\vec{\bf q}}^2$ becomes negative for negative $\bar{u}$. We obtain $q_c = \sqrt{[ ( \lambda + \mu ) |\bar{u}| ] / \kappa}$. For $\bar{u} = - 0.04$ we find $q_c \approx 0.6$ \AA$^{-1}$.
%This estimate shall be compared in the next section to an atomistic model for
%the graphene layer.

{\em Graphene on a substrate.}

The flexural modes of graphene on a substrate are modified by the coupling to the substrate. The leading effect at long wavelengths can be analyzed by considering the interaction energy per unit area between the graphene layer and the substrate, $V_{subs} ( h_{subs} )$, where $h_{subs}$ is the distance to the substrate. The dispersion relation for the flexural modes becomes $\omega_{| \vec{\bf q} |} \approx \sqrt{[ \kappa | \vec{\bf q} |^4 + V'' ( h_{eq} ) ] / \rho} = \sqrt{( \omega_{\vec{\bf q}}^0 )^2 + \omega_0^2}$, where $h_{eq}$ is the equilibrium distance. We can get an estimate of $V'' ( h_{eq} )$
\begin{align}
V'' ( h_{eq} ) &= \frac{V ( h_{eq} )}{d_0^2}
\end{align}
where $V ( h_{eq} )$ is the binding energy per unit area of graphene to the substrate, and $d_0$ is a length scale such that $d_0 \lesssim h_{eq}$. The binding energy between graphene and the substrate depends on the precise attraction mechanism \cite{SSFGNS08} between the two materials, and it is likely bound by the van der Waals interactions. A reasonable range of values is $5 - 50$ meV \AA$^{-2}$. For $d_0 \approx 2$ \AA, we find $\hbar \omega_0 \approx 1 - 4$ meV. The value of $\omega_0$ provides a cutoff in the expression for the thermal expansion, eq.~\ref{expansion}. Hence, the negative contribution of the flexural modes is reduced at temperatures such that $T \approx ( \hbar \omega_0 ) / k_B \approx 10 - 40$ K. For $k_B T \gg \hbar \omega_0$ the thermal expansion of graphene on a substrate should be similar to that of free standing graphene. At room temperature and $V ( h_{eq} ) \lesssim 50$ meV \AA$^{-2}$,  the anharmonic momentum cutoff, $q^\ast$ (see eq.~\ref{qstar} ) is such that $q^\ast \ll [ V ( h_{eq} ) / \kappa d_0^2 ]^{1/4}$, and the thermal expansion of graphene should be independent of the substrate.

{\em Bilayer graphene.}

In a discrete stack made of weakly coupled slabs we can expand the interlayer coupling assuming that the displacements vary slowly as a function of the two dimensional coordinate $\vec{\bf r}$, $u_{zz}^2 \rightarrow ( u_{n z} - u_{n+1 z} / d )^2 , u_{xz}^2 + u_{yz}^2 \rightarrow \left| ( \vec{\bf r}_{n+1} - \vec{\bf r}_{n-1} ) / 2  + \nabla_\parallel u_{n z} \right|^2$,
%\begin{align}
%u_{zz}^2 &\rightarrow \left( \frac{u_{n z} - u_{n+1 z}}{d} \right)^2 \nonumber \\
% u_{xz}^2 + u_{yz}^2 &\rightarrow \left|\frac{\vec{\bf r}_{n+1} - \vec{\bf r}_{n-1}}{d} + \nabla_\parallel u_{n z} \right|^2
%\label{discrete}
%\end{align}
where $d$ is the distance between the layers.

For two layers. an approximate expression is
\begin{widetext}
\begin{align}
{\cal E} &= \sum_{i=1,2} {\cal E}_i + {\cal E}_{int} \nonumber \\
{\cal E}_i &= \int d^2 \vec{\bf r} \left[ \frac{\lambda_{2D}}{2} \left( u_{i xx} + u_{i yy} \right)^2 + \mu_{2D} \left( u_{i xx}^2 + u_{i yy}^2 + 2 u_{i xy}^2 \right) + \frac{\kappa}{2} \left( \partial^2_{xx} u_{i z} + \partial^2_{yy} u_{i z} \right) \right] \nonumber \\
{\cal E}_{int} &= \int d^2 \vec{\bf r} \left\{ \frac{g_1}{2} \left( \frac{u_{1 z} - u_{2 z}}{d} \right)^2 +  \frac{g_2}{2}  \left[ \left( \frac{u_{ 1 x} - u_{2 x}}{d} + \frac{\partial_x u_{1 z} + \partial_x u_{2 z}}{2} \right)^2 + \left( \frac{u_{ 1 y} - u_{2 y}}{d} + \frac{\partial_y u_{1 z} + \partial_y u_{2 z}}{2} \right)^2 \right] + \right. \nonumber \\ &+ \left. \frac{g_3}{2} \left[ \frac{\partial_x \left( u_{1 x} + u_{2 x} \right) + \partial_y \left( u_{1 y} + u_{2 y} \right)}{2} \frac{\left( u_{1 z} - u_{2 z} \right)}{d} \right] \right\}
\end{align}
\end{widetext}
For an infinite three dimensional stack, the parameters $g_1$, $g_2$ and $g_3$ define a continuum model like the one in eq.~(\ref{ener}) with $c_{33} = g_1 / d , c_{13} = g_3 / d$ and $c_{44} = g_2 / d$.
%\begin{align}
%c_{33} &=\frac{g_1}{d} \nonumber \\
% c_{13} &= \frac{g_3}{d} \nonumber \\
%= c_{44} &= \frac{g_2}{d}
%\end{align}

If we assume that $g_2 = 0$, the in plane and out of plane modes are decoupled. The equations of motion for the out of plane modes are
\begin{align}
\rho_{2D} \partial^2_{tt} u_{1 z} &= - \kappa \left( \partial^2_{xx} + \partial^2_{yy} \right)^2 u_{1 z} - \frac{g_1}{d^2} \left( u_{1 z} - u_{2 z} \right) \nonumber \\
\rho_{2D} \partial^2_{tt} u_{2 z} &= - \kappa \left( \partial^2_{xx} + \partial^2_{yy} \right)^2 u_{2 z} - \frac{g_1}{d^2} \left( u_{2 z} - u_{2 z} \right)
\end{align}
where $\rho$ is the two dimensional mass density. In momentum space, we obtain two flexural modes, $\omega_+ ( \vec{\bf k}) = \sqrt{\kappa/ \rho} k^2 , \, \, \omega_- ( \vec{\bf k} ) = \sqrt{( \kappa k^4 + 2 g_1 / d^2 ) / \rho}$,
%\begin{align}
%\omega_+ \left( \vec{\bf k} \right) &= \sqrt{\frac{\kappa}{\rho}} k^2 \nonumber \\
%\omega_- \left( \vec{\bf k} \right) &= \sqrt{\frac{\kappa k^4 + 2 g_1 / d^2}{\rho}}
%\end{align}
where $k = | \vec{\bf k} |$.

 For $g_2 \ne 0$ the phonon frequencies are obtained from the diagonalization of the $6 \times 6$ matrix. It can be split into two $3 \times 3$ matrices by using the combinations $\vec{\bf r}_1 = \pm \vec{\bf r}_2 , u_{1 z} = \mp u_{2 z}$. The low energy modes are given by
 \begin{widetext}
 \begin{align}
 0 &=  {\rm Det} \left| \begin{array}{ccc}  {\scriptstyle ( \lambda_{2D} + 2 \mu_{2D} ) k_x^2 + \mu k_y  + \frac{g_2}{d^2} - \rho_{2D} \omega^2 } & {\scriptstyle ( \lambda_{2D} + \mu_{2D} ) k_x k_y } &{\scriptstyle \frac{g_2 k_x}{2 d} } \\ {\scriptstyle ( \lambda_{2D} + \mu_{2D} ) k_x k_y } & {\scriptstyle ( \lambda_{2D} + 2 \mu_{2D} ) k_y^2 + \mu_{2D} k_x^2 + \frac{g_2}{d^2}  - \rho_{2D} \omega^2 } & {\scriptstyle \frac{g_2 k_y}{2 d} } \\ {\scriptstyle \frac{g_2 k_x}{2 d} } &\scriptstyle{ \frac{g_2 k_y}{2 d} } &{\scriptstyle \frac{g_2 k^2}{4} + \kappa k^4 - \rho_{2D} \omega^2 }  \end{array} \right|
 \end{align}
 \end{widetext}

 The out of plane displacement couples to the longitudinal acoustical phonons. At low momenta we have $g_2 / d^2 \gg ( \lambda + 2 \mu ) k^2 , g_2 k^2 , \kappa k^4$, and we find
 \begin{align}
 \rho_{2D} \omega^2 &\approx \frac{g_2 k^2}{4} + \kappa k^4 - \frac{g_2^2 k^2 / (4 d^2)}{( \lambda_{2D} + 2 \mu_{2D} ) k^2 + g_2/d^2} \approx \nonumber \\ &\approx \kappa k^4 + \frac{( \lambda_{2D} + 2 \mu_{2D} ) d^2}{4} k^4 \times \nonumber \\ &\times \left[ 1 + O \left( \frac{( \lambda_{2D} + 2 \mu_{2D} ) k^2}{g_2 / d^2} \right) \right]
 \label{dispersion}
 \end{align}

The quartic term in this expression is consistent with the continuum analysis described below. The flexural modes acquire a contribution which is independent of the parameter $g_2$, and which scales with the three dimensional bulk modulus and with $d^3$, as the relation between two and three dimensional Lam\'e coefficients is $\lambda_{2D} , \mu_{2D} \propto \lambda d , \mu d$.

\begin{figure}
\begin{center}
\includegraphics[width=0.6\columnwidth]{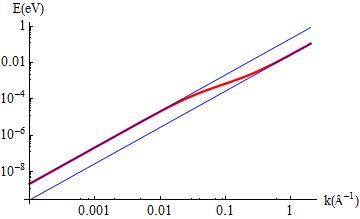}
\caption[fig]{\label{fig_flexural} Log-log plot of the dispersion of the flexural modes in a graphene bilayer using the parameters described in the text. The two straight lines correspond to the long and short wavelength limits discussed in the text.}
\end{center}
\end{figure}

For graphene, $\kappa \ll ( \lambda_{2D} + 2 \mu_{2D} ) d^2$, so that the second term dominates in eq.~(\ref{dispersion}). The bending rigidity of a bilayer should be significantly larger than that of a single layer, provided that the interlayer shear rigidity $g_2 \ne 0$. Using again $\lambda = 2$ eV \AA$^{-2}$ and $\mu = 10$ eV \AA$^{-2}$, $g_2 = 0.03$ eV and $d = 3.3$ \AA,
we find a crossover from a high to a low value of the flexural rigidity at a length $\ell = k^{-1} \approx 55$ \AA. Note that the atomistic simulations for finite-size crystallites in Ref. \onlinecite{ZLKF10} deal with a larger $k$ region giving approximately the same values for the bending rigidity (per layer) for single-layer and bilayer graphene. The dispersion of the flexural phonons is shown in Fig.~\ref{fig_flexural}.

%\begin{align}
%c_{11} &= 1060 \, \, \, {\rm GPa} \nonumber \\
%c_{12} &= 180 \, \, \,  {\rm GPa} \nonumber \\
%c_{66} &= 440 \, \, \, {\rm GPa} \nonumber \\
%c_{33} &= 36.5 \, \, \, {\rm GPa} \nonumber \\
%c_{44} &= 4.0 \, \, \, {\rm GPa} \nonumber \\
%c_{13} &= 15 \, \, \, {\rm GPa}
%$\end{align}

{\em Gr\"uneisen coefficients in a bilayer.}

By applying an in plane strain, $u$, the frequencies in eq.~(\ref{dispersion}) are reduced by
\begin{align}
\delta \left( \rho \omega^2 \right) &= - u ( \lambda_{2D} + \mu_{2D} ) k^2
\end{align}
This expression gives a Gr\"uneisen parameter
\begin{align}
\gamma_k &= - \frac{\lambda_{2D} + \mu_{2D}}{2 \left[ \kappa + ( \lambda_{2D} + 2 \mu_{2D} ) d^2 \right] k^2} \approx - \frac{\lambda_{2D} + \mu_{2D}}{2 ( \lambda_{2D} + 2 \mu_{2D} ) d^2 k^2}
\end{align}
This value is lower than the corresponding expression for single layer graphene, so that the negative expansion coefficient is reduced in a graphene bilayer. The analysis probably can be extended to graphite, although the dispersion of the out of plane modes will no longer be quadratic.

{\em Multilayered graphene. Continuum model.}

The elastic energy of a slab which is isotropic in the $x-y$ plane can be written as:

\begin{align}
{\cal E} &=  \int_{-h/2}^{h/2} d z \int d^2 \vec{\bf r} \left[ \frac{c_{12}}{2} \left( u_{xx} + u_{yy} \right)^2 + \right. \nonumber \\ &+ \frac{c_{33}}{2} u_{zz}^2 + c_{13} \left( u_{xx} + u_{yy} \right) u_{zz} + \nonumber \\ &+ \left. c_{66} \left( u_{xx}^2 + u_{yy}^2 + 2 u_{xy}^2 \right) + 2 c_{44} \left( u_{xz}^2 + u_{yz}^2 \right) \right]
\label{ener}
\end{align}
where we use te notation $c_{ij}$ for the elastic constants instead of Lam\'e coefficients.
%An alternative notation of the elastic constants is
%\begin{align}
%c_{11} &= \lambda + 2 \mu \nonumber \\
%c_{12} &= \lambda \nonumber \\
%c_{66} &= \mu \nonumber \\
%c_{33} &= \lambda' + 2 \mu' \nonumber \\
%c_{13} &= \lambda'' \nonumber \\
%c_{44} &= \mu''
%\end{align}
%\noindent
%PdA
%These elastic constants can be obtained from atomistic {\it ab-initio} calculations
%for the different systems of interest (Table \ref{tbl:cij}).

We assume that the slab is sufficiently narrow so that the stresses at the top and bottom surface do not differ much. The boundary conditions are\cite{landauEl}
\begin{align}
0 &= \sigma_{zz} =  c_{33}  u_{zz} + c_{13} \left( u_{xx} + u_{yy} \right) \nonumber \\
0 &= \sigma_{xz} = 2 c_{44} u_{xz} \nonumber \\
0 &= \sigma_{yz} = 2 c_{44} u_{yz}
\label{stresses}
\end{align}
From these equations, we obtain
\begin{align}
u_{zz} &= - \frac{c_{13} \left( u_{xx} + u_{yy} \right)}{c_{33}} \nonumber \\
u_x &= - z \partial_x u_z \nonumber \\
u_y &= - z \partial_y u_z \nonumber \\
u_{xx} + u_{yy} &= - z \left( \partial^2_{xx} u_z + \partial^2_{yy} u_z \right)
\end{align}
%The total energy can be written as
%\begin{align}
%{\cal E} &= \frac{h^3}{12} \int d^2 \vec{\bf r} \left[ \left( \frac{\lambda}{2} - \frac{{c_{13}}^2}{2  c_{33} } \right) \left( \partial^2_{xx} u_z + \partial^2_{yy} u_z \right)^2 %\right.
 %+ \nonumber \\ &+ \left. \mu \left( \left( \partial^2_{xx} u_z  \right)^2 + \left( \partial^2_{yy} u_z  \right)^2 + 2 \left( \partial^2_{xy} u_z  \right)^2 \right) \right]
 %\label{ener_slab}
%end{align}

%The kinetic energy of the slab is
%\begin{align}
%{\cal E}_{kin} &= \frac{h}{2} \int d^2 \vec{\bf r} \rho \left[ \left( \partial_t u_x \right)^2 + \left( \partial_t u_y \right)^2 + \left( \partial_t u_z \right)^2 \right]
%\label{kin}
%\end{align}
%where $\rho$ is the three dimensional mass density.

%From eq.~(\ref{ener_slab}) and eq.~(\ref{kin}),
Finally, the frequencies of the flexural modes are given by
\begin{align}
\rho \omega^2 &= \frac{h^2}{12} \left[ \lambda + 2 \mu - \frac{c_{13}^2}{c_{33}} \right] k^4
\end{align}
This expression does not depend on the value of $c_{44}$, but the value of this parameter must be different from zero, in order for eq.~(\ref{stresses}) to be valid, in agreement with the analysis carried out earlier for the bilayer.

{\em Acknowledgments.}
This work has been financed by the
MICINN, Spain, (MAT2011-26534, FIS2008-00124, FIS2011-23713, CONSOLIDER CSD2007-
00010 and CSD2007-00041), and ERC, grant 290846. MIK acknowledges financial support from FOM, the Netherlands.

%\bibliography{ripples} % Produces the bibliography via BibTeX.
% include *.bbl

\end{document}